\documentclass[preprint2]{aastex}

\usepackage{natbib}
\usepackage{amsmath}

\DeclareMathOperator\Poi{Poi}  
\DeclareMathOperator\Ga{Ga}    
\newcommand{\di}{\ensuremath{\,\text{d}}}         

\usepackage{hyperref}
\newcommand{\doi}[1]{\href{http://dx.doi.org/#1}{doi: #1}}
\newcommand{\arxiv}[1]{\href{http://arXiv.org/abs/#1}{arXiv: #1}}

\usepackage{color}

\begin{document}

 \title{Objective Bayesian analysis of ``on/off'' measurements}

 \author{Diego Casadei\altaffilmark{1}}

 \affil{School of Engineering, FHNW,
   Bahnhofstrasse 6, 5210 Windisch,
   Switzerland
 }

 \altaffiltext{1}{Visiting Scientist,
   Department of Physics and Astronomy, UCL,
   Gower Street, London WC1E 6BT, UK
 }

 \email{diego.casadei@fhnw.ch}

 \begin{abstract}
   In high-energy astrophysics, it is common practice to account for
   the background overlaid with the counts from the source of interest
   with the help of auxiliary measurements carried on by pointing
   off-source.  In this ``on/off'' measurement, one knows the number
   of photons detected while pointing to the source, the number of
   photons collected while pointing away of the source, and how to
   estimate the background counts in the source region from the flux
   observed in the auxiliary measurements.  For very faint sources,
   the number of detected photons is so low that the approximations
   which hold asymptotically are not valid.  On the other hand, the
   analytical solution exists for the Bayesian statistical inference,
   which is valid at low and high counts.  Here we illustrate the
   objective Bayesian solution based on the reference posterior and
   compare the result with the approach very recently proposed by
   \citet{knoetig2014}, discussing its most delicate points.  In
   addition, we propose to compute the significance of the excess with
   respect to the background-only expectation with a method which is
   able to account for any uncertainty on the background and is valid
   for any photon count.  This method is compared to the widely used
   significance formula by \citet{LiMa83}, which is based on
   asymptotic properties.
 \end{abstract}



 \section{Introduction}

 In a counting experiment, the detector response to a trigger signal
 is saved, whenever at least one among (possibly many) different
 conditions is satisfied.  The trigger requirements are defined in
 such a way to select interesting ``events'' and operate the detector
 in the most efficient way.  Counting experiments are widespread in
 high-energy physics and astrophysics, and sometimes have to deal with
 very low event rates.  This is the case, for example, when one tries
 to observe a very faint gamma-ray source with a space experiment,
 or when the goal is to detect the excess of counts corresponding to a
 new particle created by the collisions produced by underground
 particle accelerators.

 When only few events are collected, the asymptotic espressions which
 can be used with high count rates can not be adopted any more.
 Instead, it is of great importance to study the correct statistical
 model without simplifying assumptions which could invalidate the
 result.  For counting experiments, it is commonly assumed that the
 integer number $n \ge 0$ of observed events follows the Poisson
 distribution:
 \begin{equation*}
   \Poi(n\,|\,a) = \frac{a^n}{n!} \, e^{-a}
 \end{equation*}
 where the real value $a \ge 0$ is the Poisson parameter, which
 coincides with the expected number of events and with the variance:
 $E[n] = V[n] = a$.

 Realistic measurements always involve some degree of ``background''
 counts, due to non-interesting events which satisfy (hopefully, but
 not always, with low probability) some trigger condition.  We assume
 that the counts from the source of interest and those from the
 background are independent Poisson variables.  A well known property
 of the Poisson distribution is that the sum of two independent
 variables is again Poisson distributed, with parameter given by the
 sum of the respective expectations:
 \begin{equation}\label{eq-model}
   P(n\,|\,s,b) = \Poi(n\,|\,s+b) = \frac{(s+b)^n}{n!} \, e^{-(s+b)}
 \end{equation}
 where the expected number of events from the source $s \ge 0$ is the
 parameter of interest, and the background contribution $b \ge 0$ is
 the nuisance parameter ($n$ is integer, whereas $s$ and $b$ are real
 numbers).
 
 In high-energy astrophysics, it is common to estimate $b$ with the
 help of auxiliary measurements, obtained by pointing the detector off
 the source.\footnote{In high-energy physics, the background is
   estimate by looking at ``control regions'' in which the expected
   signal is negligible or null.  Alternatively, Monte Carlo
   simulations are used to estimate the background in the ``signal
   region''.}  In this case, it is assumed that the source of interest
 does not contribute to the observed $k$ counts, such that one has a
 simple Poisson process:
 \begin{equation}\label{eq-poisson-bkg}
   \Poi(k\,|\,B) = \frac{B^k}{k!} \, e^{-B}
 \end{equation}
 where $B \ge 0$ is the expected (background-only) photon count in the
 region off the source.  By knowing the details (like the area on the
 sky and the exposure time) of the source and off-source regions,
 it is possible to relate the expected counts from the background
 alone in the two regions: $b = \rho B$, where $\rho$ is a
 constant, assumed to be perfectly known (i.e. with negligible
 uncertainty compared to $B$).
 
 In summary, the statistical inference about this ``on/off''
 measurement makes use of the observed counts $n$ and $k$ in the
 source and off-source regions, of the known proportionality
 $\rho$ between the expected background fluxes in the two regions,
 and of the Poisson models (\ref{eq-model}) and (\ref{eq-poisson-bkg})
 for the two measurements.

 Recently, \citet{knoetig2014} (MK2014 hereafter) summarized the
 previous approaches to the on/off inference problem and proposed
 an objective Bayesian solution which consists of two different steps.
 First, it is checked whether the number $n$ of events in the source
 region is too high to be comfortably attributed to the background
 alone.  If this is the case, one rejects the ``null hypothesis''
 (background only) and claims a successful observation of the source.
 Next, the source intensity $s$ is estimated with the help of the
 auxiliary measurement.  The good point is that MK2014 finds a (rather
 complicate) analytic solution to the Bayesian inference problem, in
 terms of special (Gamma and hypergeometric) functions which are
 available in many libraries.

 The procedure proposed by MK2014 is well motivated and it aims at
 achieving very desirable goals.  However, it is not free from issues.
 The first source of possible troubles is the proposed prior [eq.~(15)
 of MK2014], which is obtained following the Jeffreys' rule in the
 bidimensional $(s,b)$ space [called
 $(\lambda_{\text{s}},\lambda_{\text{bg}})$ in MK2014].  Even though
 Jeffreys' priors have a number of desirable properties for
 1-dimensional problems, it is well known that they behave badly in
 multidimensional problems [for a recent discussion, see
 \citet{berger13}].  Here we overcome this difficulty by reducing the
 problem to a 1-dimensional (marginal) model for which an objective
 prior is known.
 
 One further complication in the procedure proposed by MK2014 is the
 comparison of the null hypothesis $H_0$ that $s=0$ (called
 ``background-only hypothesis'' here) against the alternative
 hypothesis $H_1$ that $s>0$ in the target region (the
 ``source+background hypothesis'').  In absence of additional
 information, MK2014 assigns identical (prior) probabilities to the
 two hypotheses: $P(H_0) = P(H_1) = 1/2$.  They are ``nested''
 hypotheses, in the sense that $H_0$ assumes a single value whereas
 $H_1$ allows the source strength to assume any other value in its
 domain.  The fact that the measure of the allowed domain is zero for
 $H_0$ gives troubles when using improper priors[(see for example
 \citet{bayarri08} and references therein].

 To overcome the problem, MK2014 fixes the ratio between the arbitrary
 scale factors of the source and background priors with a procedure
 relying on the assumption that, when no counts are observed both in
 the target and off-source regions, the probabilities of $H_0$ and
 $H_1$ stay the same.  This assumption is questionable, as counting
 zero events is not the same as performing no measurement.
 If the outcome is $k=0$ in the auxiliary region, we learn that the
 background in the source region is very small (possibly negligible),
 and this may imply a better sensitivity (see Appendix~\ref{sec-ul}
 for more details).  In addition, if we count $n=0$ events in the
 source region, we know that there is no background contribution in
 this outcome and also that there is no count due to the source (which
 makes it easy to get an upper bound to its intensity).  Thus,
 intuitively one may think that the absence of counts in the target
 region decreases the probability that a source is actually there,
 hence increasing $P(H_0)=1-P(H_1)$, contrarily to MK2014.
 Although this way of thinking might also be criticized, what matters
 here is that the assumption, that measuring zero counts in both
 regions does not update our degree of belief about the validity of
 both the background-only and the source+background hypotheses, is
 questionable and can not be taken as a basis for the hypothesis test.

 In this paper, we present the objective Bayesian approach based on
 the analytical solution of model (\ref{eq-model}) in the framework of
 the Bayesian reference analysis \citep{bernardo05a} obtained by
 \citet{casadei2012} (DC2012 hereafter).
 This solution is based on the ``reference prior'' corresponding to
 the model given in equation~(\ref{eq-model}), which has a solid
 formal justification and does not suffer from the problems of
 multidimensional Jeffreys' priors.  In addition, frequentist coverage
 studies have been carried on by DC2012 and show a good average
 agreement between the posterior probability and the coverage (exact
 agreement is not possible, as this is a discrete problem).

 The reference prior computed by DC2012 may be coded in any
 programming language (a C++ version is adopted here), and is also
 available in the Bayesian Analysis Toolkit \citep{BAT2009}%
 \footnote{Few lines of BAT instructions are sufficient to setup and
   solve the problem: a complete example is provided in the
   \texttt{examples/advanced/referencecounting/} directory.}.
 For the users of other analysis frameworks and/or programming
 languages, it may be useful to know that simple approximations also
 exist, which are even quicker to compute \citep{casadei2014}.

 To illustrate the application of the objective Bayesian approach of
 DC2012 and compare to MK2014, our solution will be applied in a
 simplified way (i.e.~in a single-step procedure) to the same
 Gamma-Ray Burst (GRB) data listed in table~1 of MK2014.  The marginal
 reference posterior probability density of the source strength $s$
 will be estimated directly, without comparing $H_0$ against $H_1$.
 Similarly to MK2014, the posterior distribution for $s$ will be
 summarized by providing its mode, i.e.~the most probable value or
 peak position, together with Highest Posterior Density (HPD)
 intervals, which are the narrowest intervals covering a predefined
 posterior probability.  Whenever one of such intervals is limited by
 zero at the left, its right edge automatically provides an upper
 bound to the source intensity.\footnote{It is useful to recall the
   difference between an upper limit on the intensity which one
   expects to detect with a predefined probability, characterizing the
   sensitivity of the detection technique, and an upper bound on the
   intensity inferred from the actual measurement \citep{kashyap2010}.
   Here we are only concerned with the latter.
   Appendix~\ref{sec-ul} provides details about upper limits.}
 In this case, it will be assumed that no source was detected, so that
 the alternative hypothesis of source+background is discarded.  In
 our conservative approach, when the posterior for $s$ suggests a
 non-negligible intensity we check the statistical significance of the
 excess of counts with respect to the background-only hypothesis to
 decide whether to claim a successful source detection or not.
 This procedure provides results which are essentially equivalent to
 the two-steps approach involving the comparison of two hypotheses,
 without the complications arising from the latter in the presence of
 improper priors (as it is the case for the on/off problem).


 \section{Methods}
 \label{MK2014vsDC2012}

 We have two Poisson models, characterizing a region where no source
 contribution is expected, eq.~(\ref{eq-poisson-bkg}), and the target
 region where a source may produce counts in addition to those
 coming from the background alone, eq.~(\ref{eq-model}).
 The available data are the number $n$ of photons detected when
 pointing to the source (called $N_{\text{on}}$ in MK2014), the $k$
 off-source counts (called $N_{\text{off}}$ in MK2014), and the ratio
 $\rho=b/B$ (called $\alpha$ in MK2014) between the background fluxes
 in both regions.
 One first has to estimate the background in the target region with
 the help of the auxiliary off-source measurement.  Next one considers
 the marginal model obtained by integrating over $b$ and finds the
 reference prior for $s$.  Finally, one gets the reference posterior
 for $s$ with the help of Bayes' theorem.

 The auxiliary measurement is analyzed first.  The Poisson
 distribution (\ref{eq-poisson-bkg}) allows to estimate the off-source
 background intensity $B$ by means of the Bayes' theorem:
 \begin{equation}\label{eq-off-posterior}
   p(B\,|\,k) \propto \Poi(k\,|\,B) \, \pi(B)
 \end{equation}
 (we omit the proportionality constant, as the latter can be
 determined by imposing that the integral of $p(B\,|\,k)$ be one).
 This equation expresses the posterior probability density function
 $p(B\,|\,k)$ of the off-source background intensity $B$ given the $k$
 observed counts, in terms of the likelihood function
 (\ref{eq-poisson-bkg}) and of the prior density $\pi(B)$.

 If we have some prior estimate of the background flux in the
 off-source region, it is most convenient to represent it with a Gamma
 distribution (the conjugate prior of the Poisson model):
 \begin{equation}\label{eq-gamma}
   \Ga(x\,|\,S,R) = \frac{R^S}{\Gamma(S)} \, x^{S-1} \, e^{-R x}
 \end{equation}
 with shape parameter $S>0$ and rate parameter $R>0$.  In this case,
 the posterior also belongs to the Gamma family, with new shape and
 rate parameters $S'=S+k$ and $R'=R+1$, corresponding to $k$ observed
 counts\footnote{The fact that the parameters assume new values well
   reflects the interpretation of Bayes' theorem as a way of updating
   our knowledge.}.  For example, the prior parameters $S,R$ can be
 fixed with the the method of moments, by imposing values for the
 prior expectation $E[x]=S/R$ and variance $V[x]=S/R^2$.

 In absence of prior information, it is best to adopt an objective
 prior.  The reference prior for the Poisson model coincides with
 Jeffreys' prior, which is the limiting case of a Gamma function with
 shape parameter $S=1/2$ and rate parameter $R=0$.  Hence we use here
 the (improper) Jeffreys' prior $\pi(B) = B^{-1/2}$ and find the
 (properly normalized) density which represent the solution of Bayes'
 formula (\ref{eq-off-posterior}):
 \begin{equation}\label{eq-off-posterior-2}
   p(B\,|\,k) = \Ga(B\,|\,k+\tfrac{1}{2},1)
 \end{equation}
 This is the reference posterior for the background in the off-source
 region, and the same solution is used in MK2014\footnote{The constant
   at the numerator of eq.~(14) in MK2014 is irrelevant, as an
   improper prior can have any scaling factor.  Here we took the latter to
   be one, for simplicity.  What matters is that the posterior is a
   proper density, as is the case with eq.~(\ref{eq-off-posterior-2}) above.}.

 Our goal is to estimate the source intensity $s$ in the target region.
 We use the Bayes' theorem to write the joint posterior density in
 the source region as
 \begin{equation}\label{eq-bayes-theorem}
   p(s,b\,|\,n) \propto \Poi(n\,|\,s+b) \, \pi(s) \, \pi(b)
 \end{equation}
 Later, we will integrate over $b$ to find the marginal posterior
 density $p(s\,|\,n)$.  Hence we need to find the prior $\pi(b)$ by
 translating the background estimate in the control region into a
 background estimate in the target region.
 In order to do so, we use the posterior density $p(B\,|\,k)$ from
 eq.~(\ref{eq-off-posterior-2}) to determine the prior for the
 background contribution $b$ in the source region.  With the change of
 variable $b=\rho B$ we find
 \begin{equation}\label{eq-on-bkg-prior}
   \pi(b) = \Ga(b \,|\, k+\tfrac{1}{2},\tfrac{1}{\rho})
 \end{equation}

 It is interesting to note that the expected background in the source
 region is $E[b]=\rho(k+\tfrac{1}{2})$ with the most probable value
 $\rho(k-\tfrac{1}{2})$ being the mode of $\pi(b)$ when $k\ge1$ (if
 $k=0$ the prior peaks at zero).  The commonly used maximum likelihood
 estimator $\hat{b}=\rho k$ \citep{LiMa83} is just in between the
 peak value and the expected background.  The background variance
 $V[b]=\rho^2(k+\tfrac{1}{2})$ is also very similar (but not
 identical) to the commonly used value of $\rho^2 k$.

 The next step is to write down the prior for the source strength $s$.
 We assume no prior knowledge here, hence adopt the reference prior
 calculated in DC2012.  The starting point for determining $\pi(s)$ is
 the marginal model
 \[
   P(n\,|\,s) = \int \Poi(n\,|\,s+b) \, \pi(b)
 \]
 which is in our case
 \begin{equation}\label{eq-marginal-model}
     P(n\,|\,s) = \left( \frac{1}{1+\rho} \right)^{\!k+1/2}
           e^{-s} \, f(s;n,k+\tfrac{1}{2},\tfrac{1}{\rho})
 \end{equation}
 The polynomial
 \begin{equation}\label{eq-f}
   f(x;n,c,d) = \sum_{m=0}^{n} \binom{c+m-1}{m}
                         \frac{x^{n-m}}{(n-m)! \, (1+d)^{m}}
 \end{equation}
 is a function of the real variable $x\ge0$ with integer parameter
 $n\ge0$ and real parameters $c,d>0$ whose properties are studied in
 DC2012.

 From the marginal model, DC2012 finds the Fisher's information, which
 in our case reads
 \begin{equation}\label{eq-fisher-info-2}
   \begin{split}
     I(s) &= \left( \frac{1}{1+\rho} \right)^{\!k+1/2} e^{-s}
     \; \times
     \\
     & \times \; \sum_{m=0}^{\infty}
     \frac{[f(s;m,k+\tfrac{1}{2},\tfrac{1}{\rho})]^2}
     {f(s;m+1,k+\tfrac{1}{2},\tfrac{1}{\rho})}
     - 1
   \end{split}
 \end{equation}
 The resulting reference prior is proportional to
 $\,|\,I(s)\,|\,^{1/2}$ and is improper.  Hence it is defined apart
 from a multiplicative constant.  Making use of this degree of
 freedom, the expression proposed by DC2012 is
 \begin{equation}\label{eq-ref-prior-source}
   \pi(s) = \frac{\,|\,I(s)\,|\,^{1/2} }{\,|\,I(0)\,|\,^{1/2}}
 \end{equation}
 which is a monotonically decreasing function of $s$ with maximum at
 one for $s=0$.\footnote{Incidentally, this makes it easy to compare
   it to the uniform prior, a very common (although mathematically
   ill-defined) choice with a long tradition.}
 In the limit of perfect prior background
 knowledge\footnote{$\Ga(x\,|\,c,d) \to \delta(x-b_0)$ in the limit
   $c,d\to\infty$ while keeping $b_{0}=c/d$ constant.}
 $\pi(b)=\delta(b-b_0)$, one gets Jeffreys' prior for the offset-ed
 variable $s'=s+b_0$:
 \begin{equation}\label{eq-lim-ref-prior}
   \pi(s) \longrightarrow 
          \sqrt{\frac{b_{0}}{s+b_{0}}}
          \equiv \pi_0(s)
 \end{equation}
 where $b_0 = E[b]$ is a known value.

 As it is shown in Appendix~\ref{sec-approx}, the limiting prior
 $\pi_0(s)$ can often be used in place of the more complicate
 reference prior $\pi(s)$ given in (\ref{eq-ref-prior-source}).  When
 this is not the case, a closed-form function provides an excellent
 approximation of the reference prior $\pi(s)$ as described by
 \citet{casadei2014}.  This considerably simplifies the computation,
 although it is not used in this paper.

 The marginal reference posterior for the source strength in the
 source region is finally
 \begin{equation}\label{eq-marg-post}
  p(s\,|\,n) \propto  e^{-s} \,
                      f(s;n,k+\tfrac{1}{2},\tfrac{1}{\rho})
                      \, \pi(s) \; .
 \end{equation}
 obtained after having removed the inessential constant factor
 $(1+\rho)^{-k-1/2}$ in front of the marginal
 model~(\ref{eq-marginal-model}), as the normalization of the marginal
 posterior is found by dividing by the integral from zero to infinity
 of the expression above (which is integrable for all possible values
 of the background shape and rate parameters). 

 The marginal reference posterior (\ref{eq-marg-post}) is the full
 solution of our inference problem.  Very often, it will be summarized
 by providing only few figures of merit, like the most probable value
 or the posterior expectation, plus an interval enclosing the true
 value with some predefined posterior probability.
 To complement this solution, one can check whether the counts $n$ in
 the source region are compatible with the background-only hypothesis
 by comparing $n$ with the expectation $E[b]=\rho(k+\tfrac{1}{2})$
 from a simple Poisson process in which no source is present.

 The statistical significance $z$ quantifies the deviation between
 observed counts and expected background in terms of the displacement
 from the peak of a normal distribution in units of standard
 deviations.  An excess of counts for which $z=3$ is commonly called
 ``a 3-sigma excess'' and considered a real effect (that is an
 evidence for an additional contribution on top of the events expected
 from the background-only hypothesis), although more stringent
 requirements may be preferred, like the ``5-sigma'' excess
 traditionally required in high-energy physics to claim the discovery
 of a new particle, and required by MK2014 too.\footnote{Strictly
   speaking, an excess with $z=3$ or $z=5$ leads to the rejection of
   $H_0$ with a very low probabiliy of false rejection.  It does not
   automatically imply that $H_1$ is true, as there could be more
   alternative hypotheses in competition, like e.g.~an instrumental
   effect.  However in the simplified setup in which there are only
   two alternatives, the excess is usually taken as evidence for the
   source without further discussion.  We follow this tradition here.}

 A deviation from the expected background can occur in two directions:
 as an excess of counts when $n>E[b]=\rho(k+\tfrac{1}{2})$, or as a
 deficit when $n<E[b]$.  The commonly used expression for the
 significance is eq.~(17) of \citet{LiMa83} (LM1983 hereafter).  Such
 formula gives always a positive value for $z$, while it is more
 appealing to differentiate between excess and deficit of observed
 events with respect to the expected background.  In addition,
 strictly speaking that formula is valid only asymptotically, although
 it was shown to behave well already with moderately small values of
 $n$.

 We compute the significance of the deviation from $E[b]$ with the
 inclusion of the uncertainty on the background in the source region,
 represented by the square root of the prior background variance
 $V[b]$, as described by \citet{significance} (CC2012 in the
 following)\footnote{C/C++ macros freely available on
   \url{http://svn.cern.ch/guest/psde/} based on the ROOT framework
   \citep{root09}.}
 A Poisson process with uncertain parameter, whose probability density
 is represented by a Gamma distribution, is described by the
 Poisson-Gamma mixture
 \begin{equation}
   \label{eq-Poisson-Gamma}
   \begin{split}
     P(n|c,d) &= \int_0^\infty \Poi(n\,|\,b) \, \Ga(b\,|\,c,d) \di b
     \\
     &= \frac{d^c}{\Gamma(c)} \, \frac{\Gamma(n+c)}{n! \, (1+d)^{n+c}}
   \end{split}
 \end{equation}
 In our case, we find $c,d$ with the method of moments, from $E[b] =
 c/d = \rho(k+\tfrac{1}{2})$ and $V[b] = c/d^2 =
 \rho^2(k+\tfrac{1}{2})$. 
 As expected, the result is $c=k+\tfrac{1}{2}$ and $d=1/\rho$, the
 same as in eq.~(\ref{eq-on-bkg-prior}), but
 eq.~(\ref{eq-Poisson-Gamma}) can also be used when there are other
 (e.g.~systematic) contributions to the background uncertainty in the
 target region, by applying the method of moments with the
 corresponding (larger) variance.

 For an excess, the probability $p$ to get a deviation not smaller
 than the observed one is given by the sum from $n$ to infinity of
 (\ref{eq-Poisson-Gamma}).  For a deficit, $p$ is given by the sum of
 the terms from 0 to $n$.  Next, we compute the significance $z$ by
 imposing that the integral from $z$ to infinity (excess) or from
 minus infinity to $z$ (deficit) of a standard normal distribution is
 equal to $p$. 

 This definition of $p$-value (hence of statistical significance $z$)
 is similar to the usual (frequentist) definition, with the exception
 that the Poisson-Gamma mixture (the marginal model given $H_0$) is
 used in place of the Poisson distribution, to which it reduces in
 case of negligible background uncertainty.  Hence it represents a
 sort of hybrid computation of the probability that the data deviate
 from the expectation at least as much as in the actual observation.
 The result is valid for any value of $n$, even when $n=0$, hence it
 does not rely on asymptotic properties like the LM1983 significance.
 On the other hand, MK2014 defines the $p$-value as the probability of
 the model $H_0$ given the data [eq.~(27) of MK2014].  Hence his
 significance (denoted by $S_b$) has a different meaning from the
 significance used here and by LM1983, despite from the similarity
 between the numerical values.


 \section{Results}

 We apply the methods described above on the same input data as MK2014,
 to make a detailed comparison with that solution.
 Table~\ref{tab-grb} shows gamma-ray burst (GRB) data collected by
 Fermi-LAT \citep{fermi_080825C} and VERITAS \citep{acciari11}.  Such
 GRBs were selected by MK2014 because they have low counts (otherwise
 the difference with respect to asymptotic formulae is difficult to
 notice): at least one among $n$ and $k$ is not bigger than 15 counts.

 \begin{deluxetable}{l@{\;\;}r@{\;\;}r@{\;\;}r@{\;\;}r@{\;\;}r@{\;\;}r@{\;\;}r@{\;\;}r@{\;\;}r@{\;\;}r@{\;\;}r@{\;\;}r@{\;\;}r@{\;\;}r@{\;\;}r@{\;\;}r@{\;\;}r@{\;\;}r@{\;\;}r@{\;\;}r@{\;\;}r@{\;\;}r@{\;\;}}
   \tabletypesize{\tiny}
   \tablecaption{Gamma-ray burst data from Fermi-LAT and VERITAS.}
   \label{tab-grb}
   \tablenum{1}
   \tablehead{
     \colhead{GRB} &
     \colhead{$k$} &
     \colhead{$n$} &
     \colhead{$\rho$} &
     \colhead{L99} &
     \colhead{L95} &
     \colhead{L90} &
     \colhead{L68} &
     \colhead{$E$} &
     \colhead{$M$} &
     \colhead{$P$} &
     \colhead{R68} &
     \colhead{R90} &
     \colhead{R95} &
     \colhead{R99} &
     \colhead{$V$} &
     \colhead{$S$} &
     \colhead{$K$} &
     \colhead{$\lambda_{99}$} &
     \colhead{$\lambda_{99}^{\textrm{R}}$} &
   } 

   \startdata
   070419A &  14 &  2 & 0.057 & 0.02 & 0.09 & 0.17 &  0.51 &  1.92 &  1.57 &  0.71 &  3.35 &  4.91 &  5.82 &  8.05 &  2.32 & 1.40 & 2.80 & 6.88 & 7.34 \\
   070521  & 113 &  3 & 0.057 & 0.00 & 0.00 & 0.00 &  0.00 &  1.42 &  1.02 &  0.00 &  1.67 &  3.26 &  4.20 &  6.59 &  1.83 & 1.79 & 4.58 & 6.12 & 3.52 \\
   070612B &  21 &  3 & 0.066 & 0.03 & 0.13 & 0.24 &  0.69 &  2.39 &  2.01 &  1.12 &  4.08 &  5.84 &  6.83 &  9.15 &  3.17 & 1.24 & 2.13 & 8.00 & 8.54 \\
   080310  &  23 &  3 & 0.128 & 0.00 & 0.00 & 0.00 &  0.00 &  1.87 &  1.45 &  0.00 &  2.26 &  4.07 &  5.09 &  7.46 &  2.58 & 1.49 & 3.08 & 7.16 & 7.08 \\
   080330  &  15 &  0 & 0.123 & 0.00 & 0.00 & 0.00 &  0.00 &  0.88 &  0.60 &  0.00 &  1.01 &  2.08 &  2.77 &  4.93 &  0.81 & 2.10 & 6.74 & 4.10 & 2.40 \\
   080604  &  40 &  2 & 0.063 & 0.00 & 0.00 & 0.00 &  0.00 &  1.49 &  1.11 &  0.00 &  1.77 &  3.34 &  4.25 &  6.51 &  1.85 & 1.67 & 3.99 & 6.12 & 5.66 \\
   080607  &  16 &  4 & 0.112 & 0.04 & 0.19 & 0.35 &  0.95 &  2.93 &  2.55 &  1.71 &  4.90 &  6.83 &  7.89 & 10.30 &  4.17 & 1.09 & 1.62 & 9.17 & 9.83 \\
   080825C &  19 & 15 & 0.063 & 5.95 & 7.51 & 8.39 & 10.36 & 14.27 & 13.94 & 13.28 & 18.17 & 21.27 & 22.90 & 26.29 & 15.58 & 0.50 & 0.38 & --- & ---\\
   081024A &   7 &  1 & 0.142 & 0.00 & 0.00 & 0.00 &  0.00 &  1.26 &  0.93 &  0.00 &  1.49 &  2.85 &  3.66 &  5.75 &  1.38 & 1.77 & 4.63 & 5.29 & 5.19 \\
   090418A &  16 &  3 & 0.123 & 0.02 & 0.09 & 0.17 &  0.54 &  2.15 &  1.75 &  0.57 &  3.77 &  5.52 &  6.52 &  8.94 &  2.94 & 1.34 & 2.48 & 7.64 & 8.01 \\
   090429B &   7 &  2 & 0.106 & 0.02 & 0.09 & 0.18 &  0.53 &  1.95 &  1.60 &  0.76 &  3.39 &  4.96 &  5.86 &  8.08 &  2.35 & 1.38 & 2.74 & 6.92 & 7.41 \\
   090515  &  24 &  4 & 0.126 & 0.02 & 0.09 & 0.19 &  0.51 &  2.36 &  1.93 &  0.52 &  2.88 &  6.05 &  7.13 &  9.68 &  3.54 & 1.30 & 2.27 & 8.34 & 8.66 \\
 \enddata

 \footnotesize
 \tablecomments{~Low count gamma-ray burst data where either or both
   of $n$ 
   or $k$ 
   are $\leq15$.
   All data report VERITAS measurements \citep{acciari11}, apart from
   GRB 080825C, detected by Fermi-LAT \citep{fermi_080825C}.
   The posterior HPD credible intervals are reported with 99\%, 95\%,
   90\%, 68.3\% posterior probability, together with source intensity
   expectation ($E$), median ($M$), mode ($P$), variance ($V$),
   skewness ($S$), and excess kurtosis ($K$).
   The last two columns report the 99\% upper bounds computed by
   MK2014 using his solution and the method by \citet{rolke05}.
   GRB 080825C is the only clear detection: we obtain
   $s=13.28^{+4.89}_{-2.92}$, MK2014 obtains $13.28^{+4.16}_{-3.49}$,
   and 13.7 is the official result by the Fermi-LAT collaboration.
 }
 \end{deluxetable}

\begin{figure*}[t]
   \centering
   \includegraphics[width=0.9\textwidth]{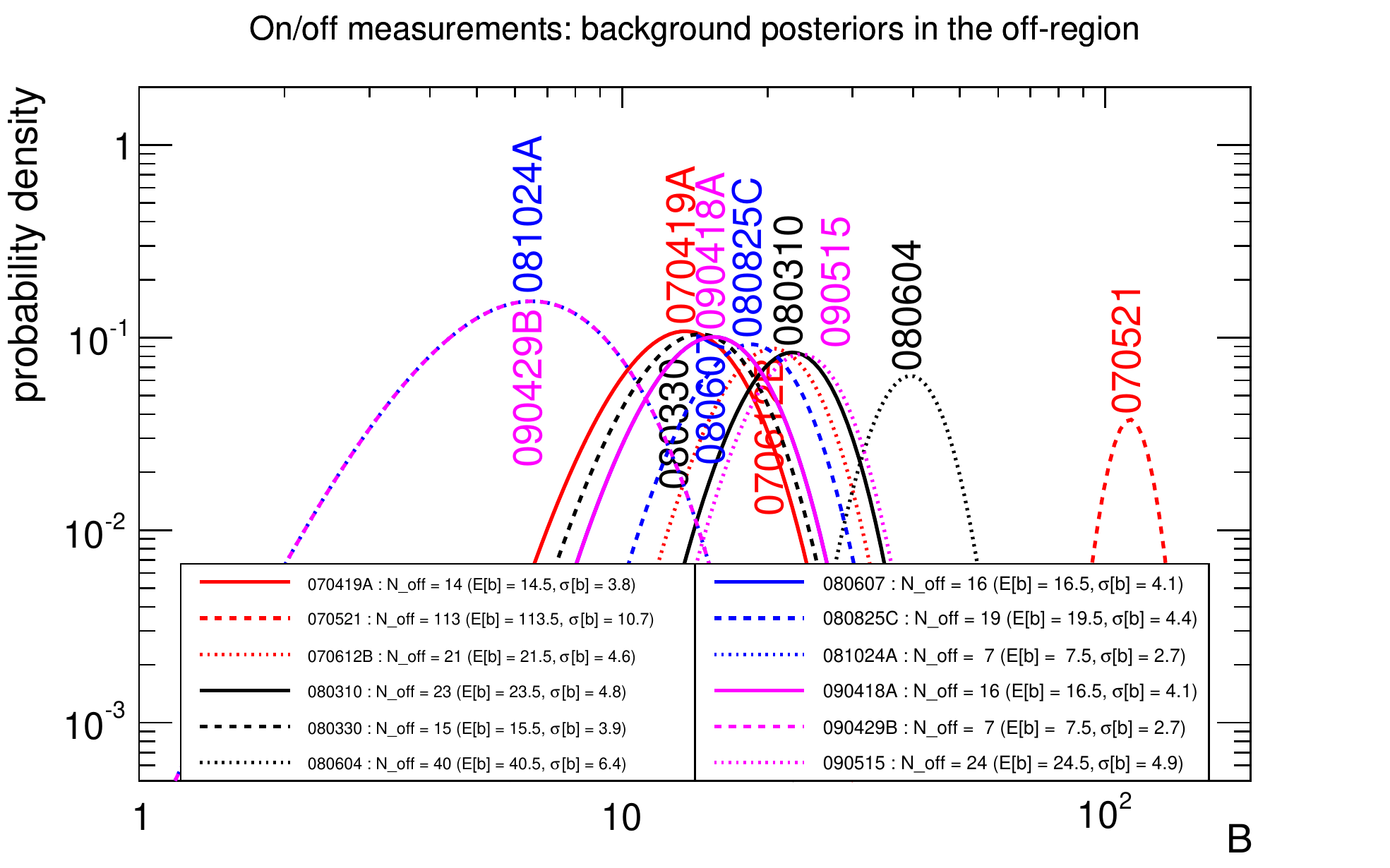}
   \caption{Reference posteriors for the background in off-source
     regions.  The GRB name is followed by the counts in the
     off-source and source regions, and by the resulting posterior
     mean and standard deviation.}
   \label{fig-bkg-off}
 \end{figure*}

 The first step is to estimate the background $B$ in the off-source
 regions with the help of eq.~(\ref{eq-off-posterior-2}).  The
 reference posterior for $B$ only depends on the off-source counts
 $k$, hence it is the same for GRBs 080607 and 090418A (both with
 $k=16$), and for GRBs 081024A and 090429B (both with $k=7$).
 Figure~\ref{fig-bkg-off} shows all the reference posteriors for the
 background $B$ in the off-source regions.  In addition to the counts
 in the off-source region, for each GRB the reference posterior mean
 and standard deviation of $B$ are reported, which may be useful
 summaries for back-of-the-envelope computations.

 Next, one finds the background prior for $b = \rho B$ in the source
 region from eq.~(\ref{eq-on-bkg-prior}).  Because the value of
 $\rho$ differs in each pair of GRBs with the same off-source
 background estimate, their background priors in the source regions
 are all distinct Gamma densities.

 Once the reference prior from eq.~(\ref{eq-ref-prior-source}) is
 computed, the final solution is provided by the (marginal) reference
 posterior for the source strength $s$ in the source region,
 eq.~(\ref{eq-marg-post}).  It is worth noticing that in all cases
 considered here, the limiting prior $\pi_0(s)$ defined in
 (\ref{eq-lim-ref-prior}) works equally well.  There would be no
 relevant change if $\pi_0(s)$ were used in place of the reference
 prior $\pi(s)$ defined in (\ref{eq-ref-prior-source}), although the
 latter was used here.

\begin{figure*}[t]
   \centering
   \includegraphics[width=0.9\textwidth]{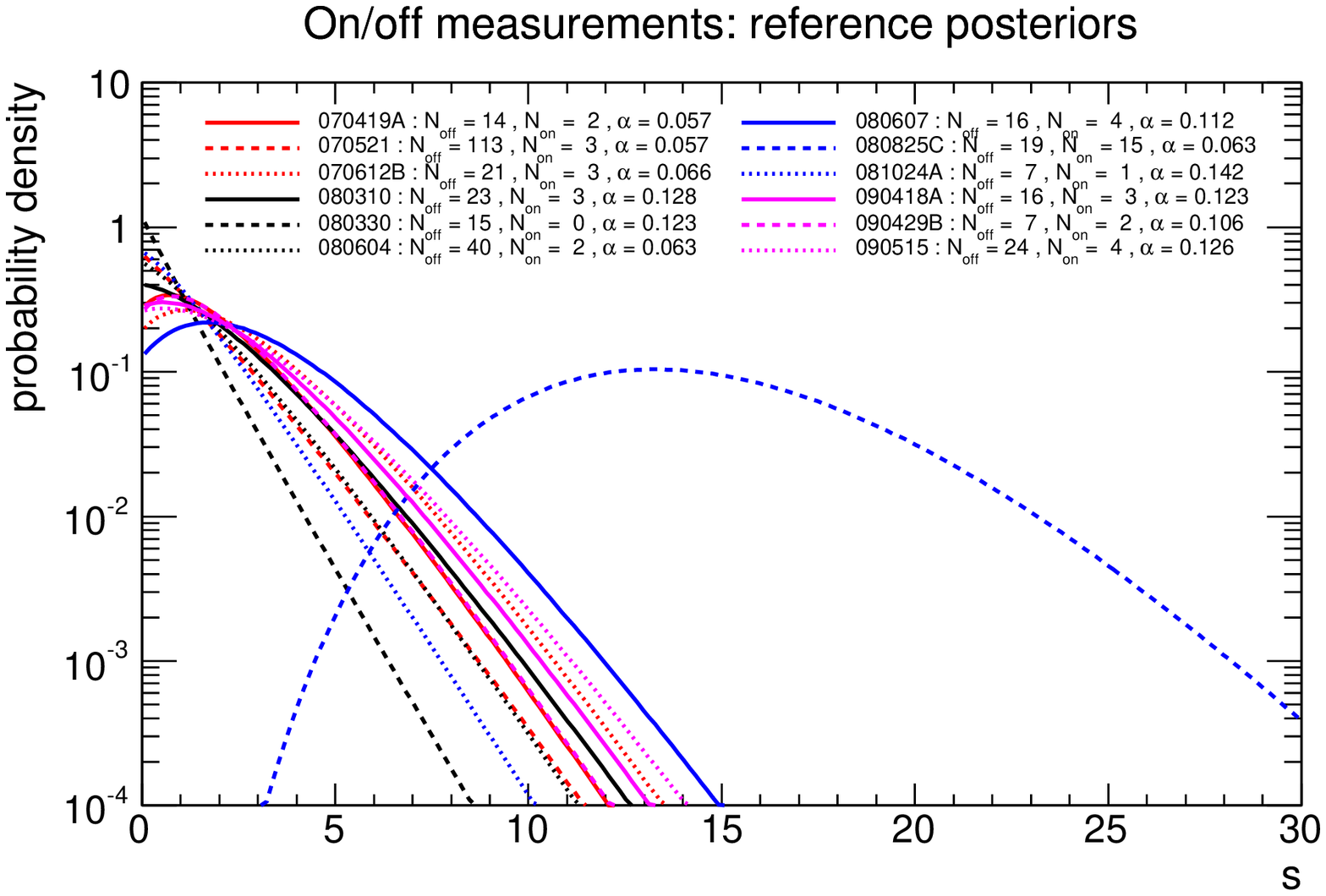}
   \caption{Marginal posteriors for GRB data.  The GRB name is
     followed by the counts in the off-source and source regions, and
     by the ratio between the background fluxes in these regions.}
   \label{fig-posteriors}
 \end{figure*}

 Figure~\ref{fig-posteriors} shows the reference posteriors for $s$
 for all GRBs listed in table~\ref{tab-grb}, where the posteriors are
 summarized by reporting the HPD credible intervals with 99\%, 95\%,
 90\%, 68.3\% posterior probability, the source intensity expectation
 ($E$), median ($M$), mode ($P$), plus variance ($V$), skewness ($S$),
 and excess kurtosis ($K$).  Two decimal places are shown in the
 table, even though they do not bring any insight on the physics,
 because the goal is to compare against MK2014 results (where only the
 99\% upper bounds are reported).

 As remarked by MK2014, the only clear detection is GRB 080825C,
 observed by Fermi-LAT \citep{fermi_080825C}: we obtain $s=
 13.28^{+4.89}_{-2.92}$ with significance $z=6.26$, whereas MK2014
 obtains $s= 13.28^{+4.16}_{-3.49}$ with significance $S_b=6.11$
 computed by mapping the posterior probability of $H_0$ onto a
 Gaussian metric.  Both Bayesian solutions find the same peak value
 for the source strength, which is only 3\% weaker than the result of
 13.7 units obtained by the Fermi-LAT collaboration, a difference ten
 times smaller than the standard deviation computed here
 ($\sqrt{V}=3.95$), hence negligible.  Our result is slightly more
 suggestive of higher source counts than MK2014, as the
 right-asymmetry of our 68.3\% credible interval is more pronounced.
 This implies that our source strength expectation (14.27 units)
 should be larger than MK2014 (where this value is not reported).
 However, this small difference is of little practical importance.
 
\begin{figure*}[t!]
   \centering
   \includegraphics[width=0.7\textwidth]{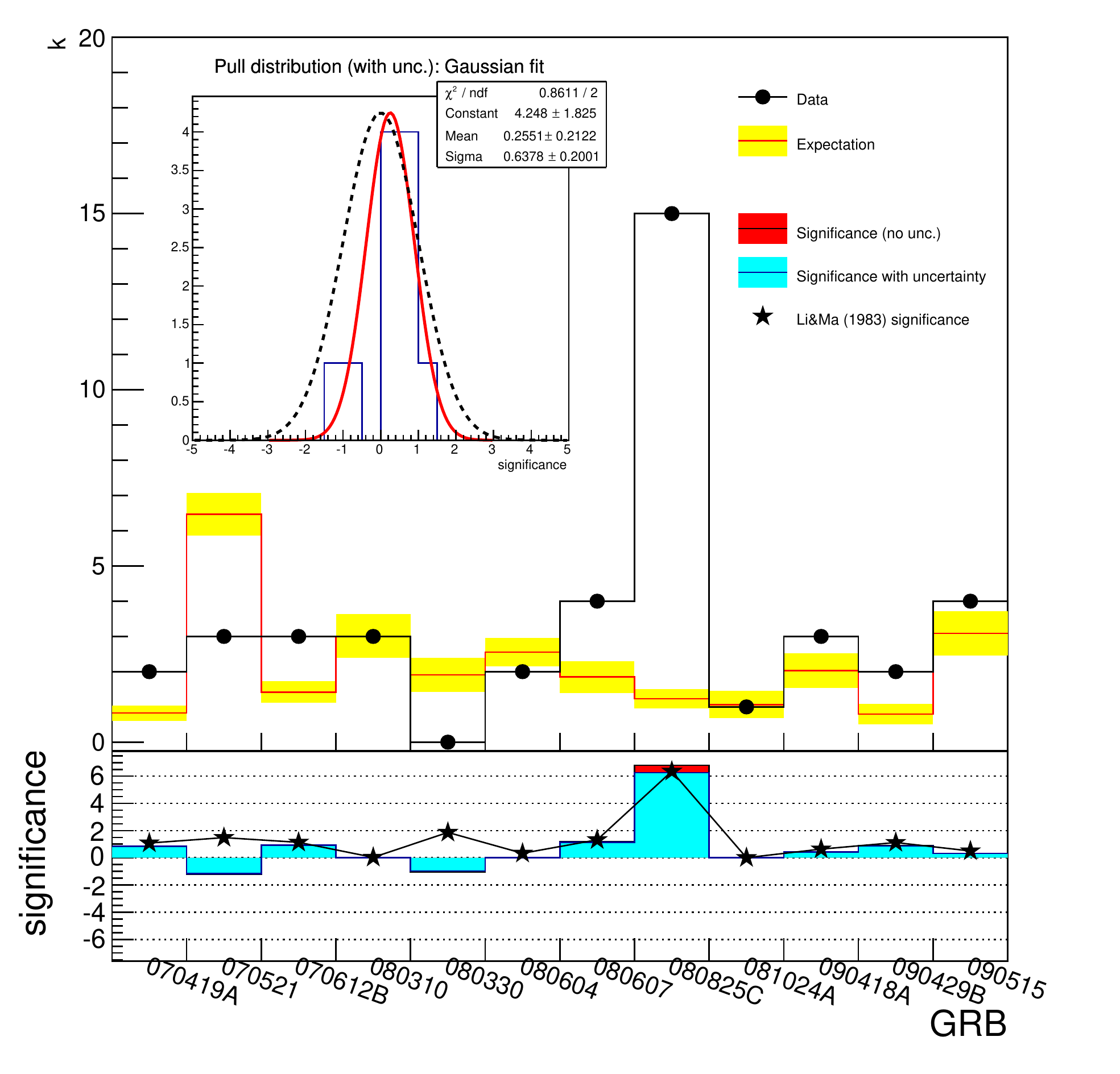}
   \caption{The observed counts (full dots) in the source region of
     each GRB are compared to the expected background and its
     uncertainty (red histogram with yellow bands representing
     fluctuations of one standard deviation in both directions).  The
     plot at the bottom shows the significance of the deviation in
     each bin, computed with (cyan histogram) and without (red
     histogram) accounting for the uncertainty on the background.  The
     black stars correspond to the significance computed accordingly
     to \citet{LiMa83}.  The inset in the top-left corner shows the
     pull distribution computed with background uncertainties, with a
     Gaussian fit (red line).  A standard normal distribution
     (centered at zero with unit standard deviation) is also shown for
     comparison (dashed black line).}
   \label{fig-significance}
 \end{figure*}

 Apart from 080825C, all other GRB data do not show any evidence for a
 detectable source, as the significance values reported in
 figure~\ref{fig-significance} confirm.  For GRBs 070521, 080310,
 080330, 080604 and 081024A, the reference posterior is monotonically
 decreasing with maximum probability density at zero: the right edges
 of their HPD intervals are all upper bounds to the source strength.
 Although the posteriors of the other GRBs are not monotonically
 decreasing functions, their peaks are so near to zero that one has in
 practice upper bounds also in these cases. 
 The significance $z$ of the deviation from the background-only
 expectation is
 0.83 for 070419A,
 0.93 for 070612B,
 1.14 for 080607,
 0.44 for 090418A,
 0.87 for 090429B, and
 0.33 for 0900515: clearly $H_0$ can not be rejected.
%
%
 In addition, their posterior 99\% HPD intervals,
 when keeping a single decimal place, all start at zero.  In
 conclusion, there is no clear evidence for some additional
 contribution in addition to the photon counts expected from the
 background alone.

 It is interesting to look at the distribution of the significance $z$
 of the deviations between observed counts and expected background, as
 this is a quick way of checking that the background-only hypothesis
 is valid for the entire set of measurements.  By definition, under
 repeated sampling the model $H_0$ will give values of $z$ which are
 normally distributed with center at zero and unit standard
 deviation. 


 A plot produced with all GRBs listed in table~\ref{tab-grb} apart
 from GRB 080825C is shown in figure~\ref{fig-significance}.  The
 values computed according to LM1983 are also shown for comparison.
 As mentioned above, they are always positive, which makes a
 difference when a deficit of events is observed.  However, this case
 is less interesting than the observation of an excess of counts, for
 which the agreement is acceptable.  The formula by LM1983
 overestimates the significance when the latter is small (by 15\% when
 $z\approx1$, increasing when $z\to0$ but decreasing when $z$
 increases), which is not a big problem in practice.  When the
 significance is high, it gives very similar results to our approach.
 For example, the detection of GRB 080825C by Fermi-LAT has
 significance $z=6.26$.\footnote{For illustration purposes only,
   figure~\ref{fig-significance} also shows than one finds $z=6.80$
   when ignoring the background uncertainty.}
 The formula by LM1983 gives a value of 6.36 while MK2014 obtains
 6.11, hence the three methods agree that the result is very
 significant.  The good agreement (1.6\%) between the standard formula
 of LM1983 and our method is connected to their very similar
 background uncertainty estimates ($\rho\sqrt{k}$ for the standard
 method of LM1983 and $\rho\sqrt{k+1/2}$ in our case).  However this
 uncertainty is purely statistical: if any additional contribution
 exists, then the formula by LM1983 is not able to account for it and
 the method by CC2012 should be used instead, as it is more general
 (the uncertainty is not assumed but is an input parameter).  Another
 possibility is to compute the posterior probability of $H_0$ given
 the data, and map it onto a Gaussian metric.  This is what is done by
 MK2014, whose result is a bit (2.4\%) smaller than, but still
 appreciably similar to ours, despite from the different meaning.

 By collecting all significance values one creates the ``pull
 distribution'', which in case of purely stochastic fluctuations
 should follow a standard normal distribution (as it is indeed the
 case when simulating a large number of pseudo-experiments).  The
 inset at the top-left corner of figure~\ref{fig-significance} shows
 that the GRB measurements --- with the exception of GRB 080825C,
 detected with more than six-sigma statistical significance and not
 shown there --- do not show any strong deviation from that
 distribution (dashed black curve), which would suggest that $H_0$
 does not all for the entire sample.  Although a Gaussian fit (red
 curve) actually confirms the preference for positive fluctuations
 which is visible in the bottom plot, it also says that the results
 are more tightly clustered than expected.  The shift of the
 barycenter is not significant, being about twice as big as the
 uncertainty on its position, confirming that the null hypothesis of
 pure background counts well describes the set of observations.


 \section{Summary and discussion}

 We have illustrated how the objective Bayesian solution to the
 inference problem for the $\Poi(s+b)$ model can be applied to the
 on/off problem.  Our solution is the marginal reference posterior
 probability density for the source strength $s$, given the measured
 counts $n$ in the source region, and the auxiliary measurement of
 background-only counts summarized by the off-source counts $k$ and
 the ratio $\rho$ between the background fluxes in the two regions.
 Based on the reference prior computed by DC2012, this solution
 appears to be more conservative (higher upper bounds) when there is
 no clear detection of additional photons with respect to the
 background-only expectation in the source region, compared to the
 posterior proposed recently by MK2014 and to the frequentist method
 based on asymptotic properties of the profile likelihood test
 statistic by \citet{rolke05}.

 The approach by MK2014 also aims at providing an objective Bayesian
 result.  Its most delicate point is the choice of the prior for the
 source region.  The choice of Jeffreys' prior in the $(s,b)$ space
 may give troubles which can be avoided if one consider the marginal
 model instead (obtained by integrating over the entire range of $b$,
 weighted by its prior).  On the other hand, the marginal model is
 1-dimensional and the corresponding reference prior is known.
 Reference priors, when available, are the recommended objective
 priors in the statistics literature, as they possess a number of
 desirable properties and are well ``calibrated'' from the frequentist
 point of view.  In 1-dimensional problems, they usually coincide with
 Jeffreys' priors, but this is not true in multidimensional problems.

 Another possible source of troubles is the hypothesis testing step in
 the method proposed by MK2014, as the improper priors used both in
 the off-source and source regions are not identical.  MK2014 proposes
 an ad-hoc procedure to overcome the problem, which is based on the
 questionable assumption that measuring $k=0$ and $n=0$ does not
 change our degree of belief about the two alternative hypotheses.  We
 propose to avoid the problem by omitting the comparison between $H_0$
 and $H_1$.  One can compute the statistical significance of the
 deviation between the observed counts and the background-only
 expectation on the basis of $H_0$ alone, as a conservative check that
 the probability of claiming a false detection is low enough.  This
 complements the estimate of the source strength $s$ provided by the
 marginal reference posterior.

 In case the hypothesis test is considered a fundamental step (which
 seems not to be the case here), it might be worth noticing that
 \citet{bernardo2011} recommends to base our decision on the
 comparison between the reference posteriors of each model by means of
 an invariant information-based loss function (the ``intrinsic
 discrepancy'').  This promises to be the best way of achieving an
 universally applicable procedure which guarantees objective
 decisions.  Unfortunately, this approach to the Bayesian hypothesis
 testing is not yet widespread in the scientific community, where most
 people continue to look at the Bayes factors (ill-defined in our
 case; see also the discussion in \citet{bayarri08}).

 The HPD intervals chosen by MK2014 cover 99\% posterior probability.
 In other words, they are the shortest 99\% credible intervals on the
 source strength $s$, given the measurements in the on/off regions.
 Even though they are not invariant under reparametrization, nor it is
 the most probable value, there is little or no discussion in the
 astrophysics community about the best choice for the parameter of
 interest: everybody just looks at $s$.  Hence here we also show the
 posterior mode with 99\% credible HPD intervals, although one might
 consider more complicate ways of summarizing the result in an
 invariant way.  The interested reader can find details about the
 ``reference posterior intrinsic loss'' function which allows one to
 find the ``intrinsic estimator'' of the parameter of interest,
 together with ``intrinsic 99\% credible regions'' for $s$, in
 \citet{bernardo05b} and \citet{bernardo2007}.

 The numerical comparison with the results obtained by MK2014 shows
 that the two methods are in decent agreement, although the use of
 Jeffreys' prior leads to narrower posterior densities than the
 marginal reference posterior.  This means that the upper bounds
 obtained by MK2014 are always tighter than those obtained here.

 The only clear case of unambiguous GRB detection by is 080825C by
 Fermi-LAT \citep{fermi_080825C}.  For this GRB, MK2014 obtains
 $s=13.28^{+4.16}_{-3.49}$ whereas the result obtained here is
 $s=13.28^{+4.89}_{-2.92}$: even though the posterior peak is at the
 same position as in MK2014, our result is slightly more suggestive of
 a higher intensity, although it well overlaps with MK2014 within the
 uncertainties.

 With all GRB data considered here, the very simple approximate
 reference prior $\pi_0 \propto (s + E[b])^{-1/2}$ provides
 practically the same result as the (more complicate) reference prior.
 This is always true when the rate parameter describing the Gamma
 prior for $b$ is large enough (in practice, it is sufficient to be
 larger than a few units), or when the shape parameter is large.  In
 our case, the shape parameter of the background prior in the source
 region is $S = k + \tfrac{1}{2}$, while the rate parameter is $R =
 \tfrac{1}{\rho}$.  The approximate reference prior $\pi_0(s)$
 differs less than 1\% from the reference prior when $R>4$ \emph{or}
 $S>40$, plus a portion of the parameters space which does not satisfy
 any of these requirements (Appendix~\ref{sec-approx}).  The first
 condition is fulfilled by all GRBs in table~\ref{tab-grb}, apart from
 GRBs 080310 ($S=23.5, R=7.8$), 081024A ($S=7.5, R=7.04$), and 090515
 ($S=24.5, R=7.9$).  However these three GRBs have shape and rate
 parameters which fall in regions of the parameters space in which
 $\pi_0(s)$ differs very little from the reference prior.  This means
 that the approximate marginal reference posterior
 \begin{equation}\label{eq-lim-ref-posterior}
   p_0(s\,|\,n) \propto \Ga(s+E[b] \,|\, n + \tfrac{1}{2}, 1)
 \end{equation}
 \citep{casadei2014}
 could have been used in place of the marginal reference posterior
 (\ref{eq-marg-post}), with a considerable simplification.

 Finally, we have noticed that the significance $z$ obtained with the
 standard asymptotic formula by \citet{LiMa83}, when it is not too
 small, well agrees with the values calculated as suggested by CC2012,
 which are correct for any value of $n$ (including the case $n=0$).
 By treating the background uncertainty as a free parameter, the more
 recent definition of $z$ is more general than the standard formula
 and should be used whenever additional sources of uncertainty exist
 beyond the pure statistical fluctuations connected with the finite
 number of photons collected while pointing off-source.


 \section{Conclusion}

 The Bayesian approach proposed by MK2014 aims at providing an
 objective solution for the on/off measurement.  Two possible sources
 of troubles with this approach are connected with the use of
 Jeffreys' prior in the bidimensional space $(s,b)$ of source and
 background intensities in the target region, and with the hypothesis
 test performed with improper priors.

 Instead of using the Jeffreys prior, a different approach exists
 which provides an objective Bayesian solution as the marginal
 reference posterior for $s$ obtained after reducing the problem to a
 1-dimensional problem.  This is done by integrating over the
 background intensity $b$ in the target region, for which an
 informative prior exists.  The reference prior for this
 1-dimensional problem is known, which ensures that the solution
 possesses all the important features of the reference posteriors,
 including invariance under reparametrization and good frequentist
 properties.  Thus, it is recommended to compute the reference
 posterior, possibly with the help of some approximation of the
 reference prior when the difference is small enough (which can be
 judged by checking the background parameters as explained in
 Appendix~\ref{sec-approx}).

 A formal two-steps procedure, in which the comparison between
 background-only $H_0$ and source+background $H_1$ hypotheses is
 performed before estimating the source intensity in the target
 region, is not strictly necessary.  A simpler approach is to estimate
 $s$ directly, and only check the significance of the deviation from
 the expectation from pure background counts in case the posterior for
 $s$ suggests a non negligible source intensity.  As the significance
 is computed with $H_0$, one avoids the complications arising when
 comparing two nested hypotheses whose parameters have improper
 priors.

 The significance $S_b$ defined by MK2014 is Bayesian, in the sense
 that it corresponds to the probability that $H_0$ is true given the
 data.  Thus, it is conceptually different from the usual definition
 of statistical significance $z$, in terms of the probability of
 obtaining data with deviations at least as large as the observed one,
 given $H_0$.  The latter is a frequentist concept, which is typically
 computed with the help of asymptotic formulae as in LM1983.  Here we
 have adopted a hybrid approach which computes the probability of
 deviations given $H_0$ with the marginal model obtained after
 integrating over the background in the target region, given by a
 Poisson-Gamma mixture.  This approach is superior to LM1983 for three
 reasons: it does not rely on asymptotics, it differentiates between
 excess and deficit, and it allows to include any uncertainty on the
 background (while LM1983 only account for the statistical uncertainty
 from the auxiliary measurement).  As expected, when the significance
 is high enough, the three definitions provide results which are
 numerically very similar.  However, our approach behaves differently
 for mildly significant results, for which LM1983 tend to overestimate
 $z$, and for all other cases (including deficits, which get assigned
 positive $z$ values by LM1983).


 \appendix

 \section{Upper limits on detectable sources}\label{sec-ul}

 \citet{kashyap2010} emphasize the important difference between the
 \emph{upper bound} on the source intensity $s$ in the target region,
 given the observed number $n$ of counts and the prior knowledge about
 the background intensity $b$ in the same region, and the \emph{upper
   limit} on the detectable sources with the chosen detection
 technique.  Upper bounds (typically at 90\%, 95\% or 99\% confidence
 level) are a way of summarizing the result of a measurement, when
 there is no significant evidence of an excess with respect to the
 counts expected from the background alone.  On the other hand, upper
 limits are connected to the detection technique and can/should be
 computed before looking at the experimental outcome.  In other words,
 upper limits are connected to the \emph{sensitivity} of the chosen
 technique, and do not depend on the counts $n$ in the source region.
 However, they depend on the estimated background in the target
 region, which in our case means that they depend on the counts $k$ in
 the off-source region, on the prior knowledge about the background
 rate $B$ in this control region, and on the known ratio $\rho=b/B$
 between the background rates in both regions.  In this paper we only
 addressed upper bounds on the measured source intensity.  Here we
 provide an algorithm to compute the upper limit accordingly to
 \citet{kashyap2010}, whose approach is generally valid for any
 detection technique, for the case of on/off measurements.

 The notation by \citet{kashyap2010} is the following.  The source and
 background rates in the target region are $\lambda_S,\lambda_B$,
 respectively, with corresponding exposure times $\tau_S,\tau_B$.  The
 ratio between the two regions is $r$, and one performs two
 measurements whose results are the counts $n_S$ and $n_B$ in the
 source and off-source regions, with
 \begin{equation}
   \label{eq-Kashyap-model}
   n_B|(\lambda_B,\tau_B,r) \sim \Poi(r \tau_B \lambda_B)
   \quad \text{and} \quad
   n_S|(\lambda_S,\lambda_B,\tau_S) \sim \Poi(\tau_S (\lambda_S+\lambda_B))
 \end{equation}
 The correspondence to our notation is the following:
 \begin{equation}
   \label{eq-notation}
   \begin{array}{lll}
     n \equiv n_S & \;  s \equiv \tau_S \lambda_S & \;  B \equiv r \tau_B \lambda_B
     \\
     k \equiv n_B & \;  b \equiv \tau_S \lambda_B & \;  \rho \equiv \tau_S/(r \tau_B)
   \end{array}
 \end{equation}

 In addition to (i) the background prior $\pi(b)$ in the source
 region, an upper limit also depends on (ii) the ``size of the test''
 $\alpha$, that is the predefined maximum tolerable probability of
 false detection (or Type I error rate), and on (iii) the predefined
 minimum ``power of the test'' $\beta$, which is connected with Type
 II errors ($\beta = 1 -$ Type II error rate), happening when one
 makes a false exclusion (i.e.~does not recognize that a source is
 present in the target region).  Common values for $\alpha$ are 0.05,
 0.003 (a ``three-sigma'' threshold), and 0.001, whereas a
 ``five-sigma'' threshold would correspond to $5.7\times10^{-7}$.  In
 addition, typically one sets $\beta=0.5$ or $\beta=0.9$ to require at
 least 50\% or 90\% detection probability, when speaking about the
 expected sensitivity of the measurement process for a given source strength.

 For the on/off measurement we choose $n$ as the test statistic and
 fix a threshold $n_\text{min} \ge 0$ such that if $n > n_\text{min}$
 we claim that a source has been detected, and if $n\le n_\text{min}$
 we declare that the measurement is compatible with the
 background-only hypothesis.  The threshold $n_\text{min}$ is chosen
 such that the probability of false detection (i.e.~of wrongly
 rejecting the null hypothesis $H_0$) does not exceed $\alpha$:
 \begin{equation}
   \label{eq-threshold}
   \alpha \ge P(n > n_\text{min} \,|\, s=0) =
   1 - \sum_{n=0}^{n_\text{min}} \int \Poi(n|b) \pi(b) \di b =
   1 - \frac{1}{(1+\rho)^{k+1/2} \, \Gamma(k+\tfrac{1}{2})}
       \sum_{n=0}^{n_\text{min}} \frac{\Gamma(n+k+\tfrac{1}{2})}%
       {n! \, (1+1/\rho)^{n}}
 \end{equation}
 where we used the Poisson-Gamma mixture from
 eq.~(\ref{eq-Poisson-Gamma}) with $c=k+\tfrac{1}{2}$ and
 $d=1/\alpha$, i.e.~we used the background prior $\pi(b)$ from
 eq.~(\ref{eq-on-bkg-prior}).  Because the threshold $n_\text{min}$
 changes at discrete steps, the actual Type I error rate will
 typically be smaller than $\alpha$.

 The expected detection probability when $s>0$ is
 \begin{equation}
   \label{eq-detection}
   \begin{split}
   \beta(s|\alpha) &= P(n > n_\text{min} \,|\, s) =
   1 - \sum_{n=0}^{n_\text{min}} \int \Poi(n|s+b) \pi(b) \di b =
   1 - \frac{e^{-s}}{(1+\rho)^{\!k+1/2}}
        \sum_{n=0}^{n_\text{min}} f(s;n,k+\tfrac{1}{2},\tfrac{1}{\rho})
   \\
    &= 1 - \frac{e^{-s}}{(1+\rho)^{\!k+1/2} \, \Gamma(k+\tfrac{1}{2})}
       \sum_{n=0}^{n_\text{min}} \sum_{m=0}^{n} \frac{\Gamma(m+k+\tfrac{1}{2}) \, s^{n-m}}%
       {m! \, (n-m)! \, (1+1/\rho)^{m}}
   \end{split}
 \end{equation}
 obtained after inserting the polynomial (\ref{eq-f}) into the
 marginal model (\ref{eq-marginal-model}).  For $s\to0$
 eq.~(\ref{eq-detection}) gives the last expression of
 (\ref{eq-threshold}), because only the term with $m=n$ survives in
 the second sum.  Thus $\beta(0|\alpha) \le \alpha$, which means that
 one must find a compromise, while aiming at a low probability of
 claiming a fake source and a high probability to detect a true but
 faint source.
 
 Now that all ingredients are available, we can formulate the
 algorithm for computing the upper limit.
 Once the size and power of the test have been chosen (say
 $\alpha=0.003$ and $\beta=0.5$), one has to estimate the background
 in the target region, where the prior $\pi(b)$ shall be chosen as a
 Gamma density whose parameters are determined by the method of
 moments, starting from the background expectation and variance.  In
 the on/off measurement, we use the control region to find the
 background in the target region, given by
 eq.~(\ref{eq-on-bkg-prior}).  Next, we define a threshold for
 detecting the source, in terms of the minimum number $n_\text{min}$
 of counts which ensures that our Type I error rate does not exceed
 $\alpha$, following the inequality (\ref{eq-threshold}).  Finally,
 one takes as the upper limit the smallest value of $s$ for which
 $\beta(s|\alpha)\ge0.5$, where $\beta(s|\alpha)$ is computed in
 eq.~(\ref{eq-detection}).


 \section{Approximate forms for the reference prior}\label{sec-approx}

 Here we report useful approximations to the reference prior $\pi(s)$
 defined in eq.~(\ref{eq-ref-prior-source}), from \citet{casadei2014}.
 A movie comparing the reference prior with these approximations and
 with the flat prior is available on
 \url{https://www.youtube.com/watch?v=vqUnRrwinHc}, clearly showing
 when different approximations should be used.

 The limiting form $\pi_0(s)$ of the reference prior when there is
 certain knowledge of the background in the source region is given by
 eq.~(\ref{eq-lim-ref-prior}).  The limit of perfect knowledge is
 approached by increasing values of the shape parameter, as the
 relative uncertainty on the background in the source region is
 $\sqrt{V[b]}/E[b] = 1 / \sqrt{S}$.  However, it turns out that even
 at small values of $S$ there are cases in which $\pi_0(s)$ provides a
 very good approximation to $\pi(s)$.

 In order to quantify the deviation from $\pi(s)$, their relative RMS
 difference has been computed on the signal range $0 \le s \le 70$,
 by dividing the distance
 \begin{equation}
   d_2 = \left( \int_{0}^{70} [\pi_0(s) - \pi(s)]^2 \di s \right)^{\!1/2}
 \end{equation}
 by the integral of the reference prior over the same range.

\begin{figure}[t!]
   \centering
   \includegraphics[width=0.6\textwidth]{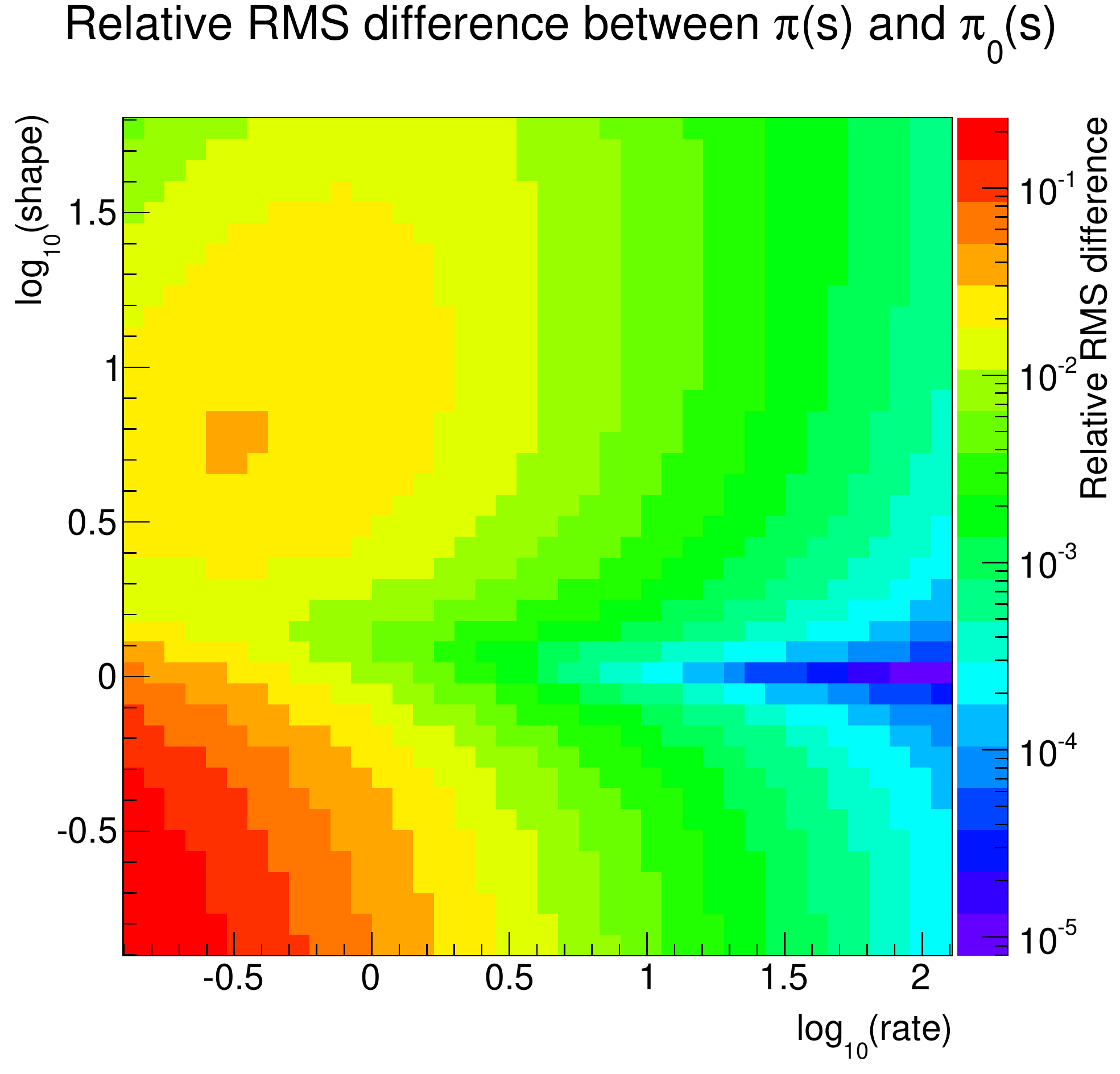}
   \caption{Relative RMS difference between $\pi(s)$ and $\pi_0(s)$ as
     a function of the shape and rate parameters of the background
     prior in the source region \citep{casadei2014}.}
   \label{fig-f0}
 \end{figure}

 For most practical purposes, a relative RMS difference below 1\% is
 acceptable, as this is the order of magnitude of the maximum change
 in the posterior in the limit of very few or zero observed counts.
 For increasing $n$, the changes of the posterior become smaller and
 smaller.  Figure~\ref{fig-f0} shows that the $\pi_0(s)$ is
 satisfactory (differing from $\pi(s)$ by less than 1\%) when the
 shape parameter is larger than 40 \emph{or} the rate parameter is
 larger than 4, and in some case even for lower values.

 It should be emphasized that the threshold at 1\% chosen here is
 arbitrary and quite conservative.  In most applications larger
 deviations can be acceptable, as the posteriors will quickly become
 indistinguishable for increasing number $n$ of observed counts.  In
 addition, the common practice is to summarize the posterior by
 providing one value (e.g.~the expectation or the mode) and some
 estimate of its uncertainty (e.g.~the shortest interval covering
 68.3\% posterior probability), by rounding the values to the minimum
 meaningful number of digits.  Often, this summary is quite robust
 compared to relative RMS differences of several percent.


 %


\begin{thebibliography}{99}%

 \bibitem[Abdo et al.(2009)]{fermi_080825C}
   A.A.\ Abdo et al.,
   \emph{FERMI observations of high-energy gamma-ray emission from GRB
     080825C},
   \apj 707 (2009) 580,
   \doi{10.1088/0004-637X/707/1/580}.

 \bibitem[Acciari et al.(2011)]{acciari11}
   V.A.\ Acciari et al.,
   \emph{VERITAS observations of gamma-ray bursts detected by SWIFT},
   \apj 743 (2011) 62,
   \doi{10.1088/0004-637X/743/1/62}

 \bibitem[Antcheva et al.(2009)]{root09}
   I.\ Antcheva et al.,
   \emph{ROOT - A C++ framework for petabyte data storage, statistical
     analysis and visualization},
   Computer Physics Communications 180 (2009) 2499,
   \doi{10.1016/j.cpc.2009.08.005}.

 \bibitem[Bayarri et al.(2008)]{bayarri08}
   M.J.\ Bayarri, J.O.\ Berger \& G.S.\ Datta,
   \emph{Objective Bayes testing of Poisson versus inflated Poisson models},
   IMS Collections 3 (2008) 105,
   \doi{10.1214/074921708000000093}.
  
 \bibitem[Berger et al.(2013)]{berger13}
   J.O.\ Berger, J.M.\ Bernardo \& D.\ Sun,
   \emph{Overall objective priors},
   Tech.\ Rep.\ (Duke University, USA), 2013
   \url{http://www.uv.es/~bernardo/BBS2013.pdf}.

 \bibitem[Bernardo(2005a)]{bernardo05a}
   J.M.\ Bernardo,
   \emph{Reference analysis},
   Handbook of Statistics 25 (D.K.\ Dey and C.R.\ Rao eds.). Amsterdam:
   Elsevier (2005) 17--90. 

 \bibitem[Bernardo(2005b)]{bernardo05b}
   J.M.\ Bernardo,
   \emph{Intrinsic credible regions: An objective Bayesian approach to interval
     estimation},
   Test 14 (2005) 317,
   \doi{10.1007/BF02595408}.

 \bibitem[Bernardo(2007)]{bernardo2007}
   J.M.\ Bernardo,
   \emph{Ojective Bayesian point and region estimation in location-scale models},
   Sort 31 (2007) 3-44.

 \bibitem[Bernardo(2011)]{bernardo2011}
   J.M.\ Bernardo,
   \emph{Integrated Objective Bayesian Estimation and Hypothesis Testing},
   Bayesian Statistics 9 (2001) 1,
   \doi{10.1093/acprof:oso/9780199694587.003.0001}.
  
 \bibitem[Casadei(2012)]{casadei2012}
   D.\ Casadei,
   \emph{Reference analysis of the signal + background model in
     counting experiments},
   JINST 7 (2012) P01012, 
   \doi{10.1088/1748-0221/7/01/P01012},
   \arxiv{1108.4270} [DC2012].

 \bibitem[Casadei(2014)]{casadei2014}
   D.\ Casadei,
   \emph{Reference analysis of the signal + background model in
     counting experiments II. Approximate reference prior},
   \arxiv{1407.5893}, 2014\_JINST\_9\_T10006, \doi{10.1088/1748-0221/9/10/T10006}.

 \bibitem[Caldwell et al.(2009)]{BAT2009}
   A.\ Caldwell, D.\ Kollar \& K.\ Kr\"oninger,
   \emph{BAT - The Bayesian Analysis Toolkit},
   Computer Physics Communications 180 (2009) 2197
   \doi{10.1016/j.cpc.2009.06.026}, \arxiv{0808.2552}.

 \bibitem[Choudalakis\&Casadei(2012)]{significance}
   G.\ Choudalakis \& D.\ Casadei,
   \emph{Plotting the Differences Between Data and Expectation},
   Eur.\ Phys.\ J.\ Plus 127 (2012) 25,
   \doi{10.1140/epjp/i2012-12025-y},
   \arxiv{1111.2062} [CC2012].

 \bibitem[Kashyap et al.(2010)]{kashyap2010}
   V.L.\ Kashyap et al.,
   \emph{On Computing Upper Limits to Source Intensities},
   Astrophys.\ J.\ 719 (2010) 900,
   \doi{10.1088/0004-637X/719/1/900}

 \bibitem[Knoetig(2014)]{knoetig2014}
   M.L.\ Knoetig,
   \emph{Signal discovery, limits, and uncertainties with sparse On/Off
     measurements: an objective Bayesian analysis},
   Astrophys.\ J.\ 790 (2014) 106,
   \doi{10.1088/0004-637X/790/2/106},
   \arxiv{1406.2922} [MK2014].

 \bibitem[Li\&Ma(1983)]{LiMa83}
   T.-P.\ Li \& Y.-Q.\ Ma,
   Astrophys.\ J.\ 272 (1983) 317,
   \doi{10.1086/161295} [LM1983].

 \bibitem[Rolke et al.(2005)]{rolke05}
   W.A.\ Rolke, A.M.\ L\'{o}pez \& J.\ Conrad,
   Nucl.\ Instrum.\ Methods A 551 (2005)  493,
   \doi{10.1016/j.nima.2005.05.068},
   \arxiv{physics/0403059}

\end{thebibliography}
\end{document}